# IMPROVE PERFORMANCE OF TCP NEW RENO OVER MOBILE AD-HOC NETWORK USING ABRA

Dhananjay Bisen[1] and Sanjeev Sharma[2]

[1]M.Tech, School Of Information Technology, RGPV, BHOPAL, INDIA

[1]`bisen.it2007@gmail.com`

[2]Reader & Head, School Of Information Technology, RGPV, BHOPAL, INDIA

[2]`sanjeev@rgtu.net`

## ABSTRACT

*In a mobile ad hoc network, temporary link failures and route changes occur frequently. With the assumption that all packet losses are due to congestion, TCP performs poorly in such an environment. There are many versions of TCP which modified time to time as per need. In this paper modifications introduced on TCP New Reno over mobile ad-hoc networks using calculation of New Retransmission Time out (RTO), to improve performance in term of congestion control. To calculate New RTO, adaptive backoff response approach (ABRA) in TCP New Reno was applied which suggest ABRA New Reno. It utilizes an ABRA by which congestion window and slow start threshold values were decreased whenever an acknowledgement is received and new backoff value calculate from smoothed round trip time. Evaluation based on comparative study of ABRA New Reno with other TCP Variants like New Reno and Reno was done using realistic parameters like TCP Packet Received, Packet Drop, Packets Retransmitted, Throughput, and Packet Delivery Ratio calculated by varying attributes of Node Speed, Number of Nodes and Pause Time. Implementation and simulations were performed in QualNet 4.0 simulator.*

## 1. INTRODUCTION

Mobile ad hoc networks (MANETs) [1] are collections of mobile nodes, dynamically forming a temporary network without centralized administration. These nodes can be arbitrarily located and are free to move randomly at any given time, thus allowing network topology and interconnections between nodes to change rapidly and unpredictably. There has been significant research activity over the past 10 year into performance of such networks with the view to develop more efficient and robust TCP variants. Transmission control protocol [2] provides reliability, end-to-end congestion control mechanism, byte stream transport mechanism, flow control, and congestion control. Comparing to wire networks, there are many different characteristics in wireless environments, which makes TCP congestion control mechanism is not directly suitable for wireless networks and many improved TCP congestion control mechanisms [3] have been presented. However, TCP in its present form is not well suited for mobile ad hoc networks. In addition to all links being wireless, frequent route failures due to mobility can cause serious problems to TCP as well. Route failures can cause packet drops at the intermediate nodes, which will be misinterpreted as congestion loss.

When a route failure occurs for a period of time greater than retransmission timer value [4] [8], TCP understand this as congestion which means decreasing both the congestion window (CWND) and slow start threshold (SSThr). Subsequently it retransmits the first unacknowledged packet and executes back-off by doubling the value of retransmission timer. Under multiple successive back-offs, the value of the retransmission timer is too long, However during the long retransmission period, the route may come back but TCP will not try to retransmit the first unacknowledged packet until the retransmission timer expires. The ideal solution for the route failure problem is to freeze its state as soon as the route breaks and resume as soon as a new

International Journal of Wireless & Mobile Networks (IJWMN) Vol. 3, No. 2, April 2011

route is found. Most of research are concentrating on congestion, corruption control and improve retransmission time out condition, improving performance metrics and security threats of protocol. Hence this thrust area of mobile ad-hoc network become the choice of interest for us. This paper is organized in such a manner that section 2 describes Congestion Control Algorithm followed by TCP New Reno, Adaptive Backoff Response Approach and new retransmission time out in section 3, 4 and 5 respectively. In section 6, simulation environment and methods is described with simulation results and analysis of the performance in 7.

## 2. CONGESTION CONTROL ALGORITHM

TCP is a transport layer protocol used by applications that require guaranteed delivery and reliable transmission of packets. It is a sliding window protocol that provides handling for both timeouts and retransmissions. Another function is Congestion control [6] [7]. It is a method used for monitoring the process of regulating the total amount of data entering the network so as to keep traffic levels at an acceptable value, whereas flow control controls the per-flow traffic such that the receiver capacity is not exceed. Congestion control mostly applies to packet-switching network. A wide variety of approaches have been proposed, however the "objective is to maintain the number of packets within the network below the level at which performance falls off dramatically."

The TCP sender starts the session with a congestion window value of one maximum segment size (MSS) it is a parameter of the TCP protocol that specifies the largest amount of data, specified in bytes, that a computer can receive in a single, unfregmented piece. The window size determines the number of bytes of data that can be sent before an acknowledgement from the receiver is necessary. This doubling of the congestion window with every successful acknowledgment of all the segments in the current congestion window is called slow-start or exponential start show in figure 1 and it continues until the congestion window reaches the slow-start threshold.

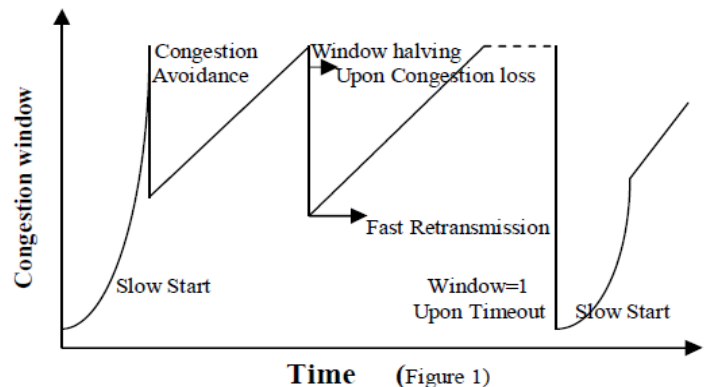

Once it reaches the slow-start threshold, it grows linearly, adding one MSS to the congestion window on every ACK received. This linear growth, which continues until the congestion window reaches the receiver window, is called congestion avoidance, [5] show in figure 1 as it tries to avoid increasing the congestion window exponentially, which will surely worsen the congestion in the network.

TCP updates the RTO period with the current round-trip delay calculated on the arrival of every ACK packet. If the ACK packet does not arrive within the RTO period, then it assumes that the packet is lost due to the congestion in the network and if TCP sender receives three consecutive duplicate ACKs (DUPACKs). Upon reception of three DUPACKs, the TCP sender retransmits the oldest unacknowledged segment. This is called the fast retransmit scheme show in figure 1. The TCP sender does the following during congestion: Reduces the slow-start threshold to half the current congestion window, Resets the congestion window size to one MSS,

Activates the slow-start algorithm, and Resets the RTO with an exponential back-off value which doubles with every subsequent retransmission.

A variety of TCP approaches have been proposed like TCP Tahoe [6], TCP Reno [9], TCP with selective ACK (SACK) [9]. There are used various congestion control algorithms slow start threshold, congestion avoidance, congestion detection, fast recovery, fast retransmission. This is for detect losses and congestion. That is, TCP uses timeout and duplicate ACKs to detect loss, and changes in round-trip times to detect congestion.

## 3. TCP NEW RENO

The experimental version of TCP Reno is known as TCP New Reno [5]. It is slightly different than TCP Reno in fast recovery algorithm. New Reno is more competent than Reno when multiple packets losses occur. New Reno and Reno both correspond to go through fast retransmit when multiple duplicate packets received, but it does not come out from fast recovery phase until all outstanding data was not acknowledged . It implies that in New Reno, partial ACK do not take TCP out of fast recovery but they are treated as an indicator that the packet in the sequence space has been lost, and should be retransmitted. Therefore, when multiple packets are lost from a single window of data, at this time New Reno can improve without retransmission time out. The retransmitting rate is one packet loss per round trip time until all of the lost packets from that window have been transmitted. It exist in fast recovery till all the data is injected into network, and still waiting for an acknowledgement that fast recovery was initiated.

The critical issue in TCP New Reno [12] is that it is capable of handling multiple packet losses in a single window. It is limited to detecting and resending only one packet loss per round - trip-time. This insufficiency becomes more distinct as the delay-bandwidth becomes greater. However, still there are situations when stalls can occur if packets are lost in successive windows, like all of the previous versions of TCP New Reno which infer that all lost packets are due to congestion and it may therefore unnecessarily cut the congestion window size when errors occur. There are [12] some steps of congestion control for New Reno transmission control protocol.

**Step 1:** Initially
    0<cwnd<= min (4*mss, max (2*mss, 4380 bytes))
    SS_threshold = max (cwnd/2, 2*MSS)
**Step 2:** Slow Start Algorithm (Exponential Increases)
    If (receive acks && cwnd< ss_threshold)
    cwnd = cwnd+1;
**Step 3:** Congestion Avoidance Algorithm (Additive Increase)
    If (receive ACKs) {
    If (cwnd > ss_threshold)
       cwnd = cwnd + segsize * segsize / cwnd;
    Else
      cwnd = cwnd + 1;   }
**Step 4:** Congestion Detection Algorithm (Multiplicative Decrease): Fast Retransmission and Fast Recovery
   If (congestion) {
   If (Receive same Acks 3 time or retransmission time out) {
   SS_threshold = cwnd/2;
   If (Retransmission time out) {
           cwnd = initial;
   Exit and call Slow Start step;
   Else /* Receive same Acks 3 time*/
           cwnd = SS_threshold;
   Exit and call congestion avoidance step;    }}}

## 4. ADAPTIVE BACKOFF RESPONSE APPROACH (ABRA)

In the event of a retransmit timeout, TCP retransmits the oldest unacknowledged packet and doubles the retransmit timeout interval (RTO). This process is repeated until an ACK for the retransmitted packet has been received. So retransmission timeout interval may be very long although the route may have been re-established some time ago. This leads to a wasted time. The wasted time can be used to send packets. We try to make use of this wasted time by making the retransmission timeout interval depends on a smoothed round trip time (SRTT), which is a weighted average of measured retransmitted timeout. The ABRA [10] depends on saving the values of congestion window, slow start threshold and smoothed round trip time, when the retransmission timer expires. Here instead of multiplying RTO interval by two each time, we multiply it by a value called backoff $_{new}$ between one and two depending on the last_srtt, which is a weighted average of last measured retransmitted timeout.. The backoff $_{new}$ and the new RTO value ($RTO_{new}$) are computed as follows:

$$Backoff_{new} = 1 + \frac{(last\_srtt - min\_srtt)}{(max\_srtt - min\_srtt)}$$

$$RTO_{new} = Backoff_{new} * RTO_{current}$$

where $RTO_{current}$ is the current RTO value, min_srtt is the minimum smoothed round trip time seen so far, and max_srtt is the maximum smoothed round trip time seen so far. We initialize [10] min_srtt and max_srtt with the values 0.1 and 0.6 seconds respectively choosing initialization values of these two variables.

## 5. NEW RETRANSMISSION TIMEOUT CALCULATION

There are steps of calculation of new retransmitted time out using Adaptive Backoff Response approach for New Reno transmission control protocol.

**Step 1:** Calculate Round Trip Time (RTT):- the measured RTT for a segment is the time required for the segment to reach the destination and be acknowledged.
**Step 2:** Calculate Smoothed Round Trip Time (SRTT) [8]: which is a weighted average of RTT $_{measured}$ and the previous SRTT as shown below:
For first measurement
*$SRTT_{previous} = RTT_{measured}$*
From second measurement the calculus becomes
*$SRTT = (1-a) SRTT_{previous} + a . RTT_{measured}$*
    *a = 1/8*
**Step 3:** Calculate RTT (Deviation): RTTD [8]
For first measurement
*$RTTD_{previous} = RTT_{measured}/2$*
From second measurement the calculus becomes
*$RTTD = (1-b) RTTD_{previous} + b. (SRTT - RTT_{measured})$*
    *b = 1/4*
**Step 4:** Calculate Retransmission Time out (RTO): The value of RTO is based on the smoothed round trip time and its deviation.
For first measurement
*$RTO_{current} = SRTT + 4.RTTD$*
Calculate New RTO with using SRTT; apply Adaptive back off response approach.
**Step 5:**(Proposed Step) for second measurement**,** first calculate New Back-off value and multiply that Value to RTO (previous RTO value).

$$Backoff_{new} = 1 + \frac{(last\_srtt - min\_srtt)}{(max\_srtt - min\_srtt)}$$

$$RTO_{new} = Backoff_{new} * RTO_{current}$$

Constant a and b in step 2 and step 3 have been selected on basis of the literature where a=1/8 and b=1/4 is mentioned in reference [8] as well as value of a and b are implementation dependent but under normal circumstances they are set to be 1/8 and 1/4 respectively.

## 6. SIMULATION ENVIRONMENT

All the simulation work is perform in QualNet wireless network simulator version 4.0 [13]. Initially number of nodes are 50, Simulation time was taken 200 seconds and seed as 1. All the scenarios have been designed in 1500m x 1500m area. Mobility model used is Random Way Point [14] (RWP). The mobility model is designed to describe the movement pattern of mobile users, and how their location, velocity and acceleration change over time. In random-based mobility simulation models, the mobile nodes move randomly and freely without restrictions. To be more specific, the destination, speed and direction are all chosen randomly and independently of other nodes. For simulation, environmental surrounding selected are speed, Pause time and no. of mobile nodes. Speed and pause time are between the range of 5-30 m/s and sec. respectively while no. of nodes are between the range of 10-50. "Pause time is a time in which all nodes in network are motionless but transmission in continued". All the simulation work were carried out using TCP variants (Reno, New Reno, ABRA New Reno) with AODV routing protocol .Network traffic is provided by File Transfer Protocol (FTP) application. File Transfer Protocol represents the File Transfer Protocol server and client.

## 7. RESULTS AND ANALYSIS

The simulation for ABRA New Reno is based on simulation time, number of node, area of network, pause time, routing protocols, and speed of node. In experimental methodologies variation in one of the parameter including no. of node, speed of node and the pause time was done each time and their effect on performance of different TCP protocols were determined while rest of all other parameters like simulation time, area of network, pause time, routing protocols, and speed of node kept constant. Effects of simulation studies on performance of ABRA New Reno and other TCP protocols under experimental conditions mentioned above were represented graphically.

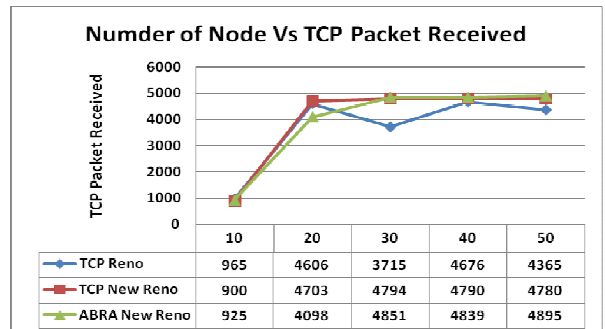

Figure 2

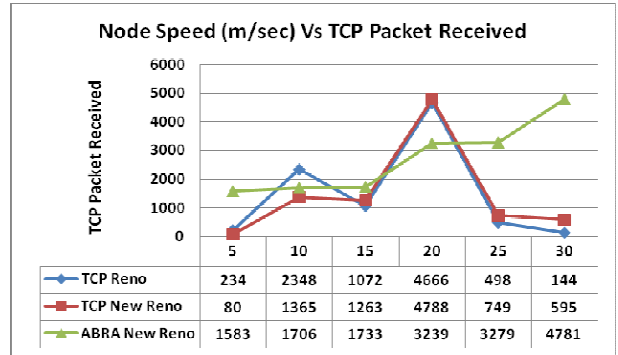

Figure 3

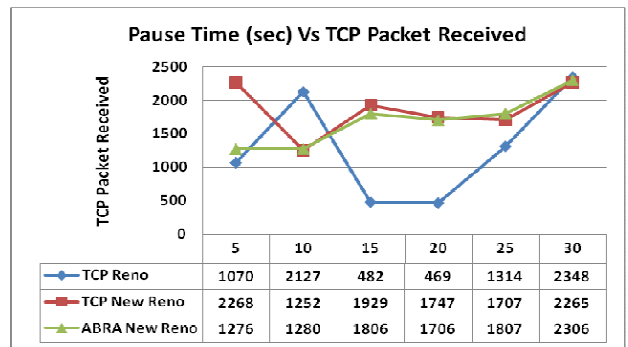

Figure 4

Simulation results in fig. 2-4, it is observed that under less no. of nodes, node speed and pause time sender does not find proper path, hence, receiver is far from source and no. of mobile nodes are least between sender and receiver, hence, time out will crop up, and therefore minimum packet received. But upon increments in variants like no. of nodes, node speed and Pause time of the network then sender will receive different paths between senders and receiver easily, hence maximum packets were received.

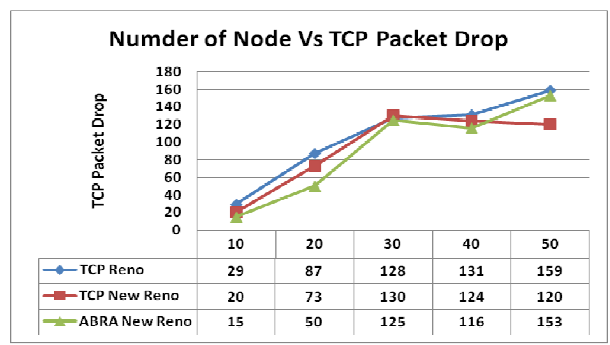

Figure 5

International Journal of Wireless & Mobile Networks (IJWMN) Vol. 3, No. 2, April 2011

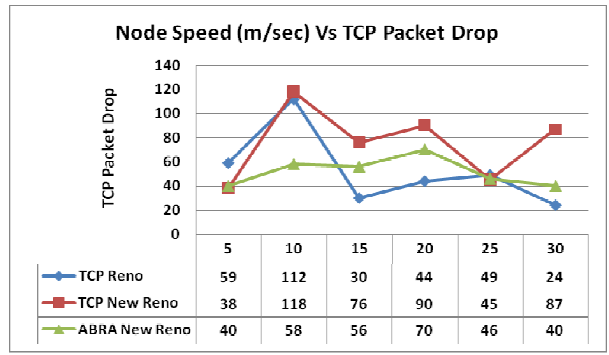

Figure 6

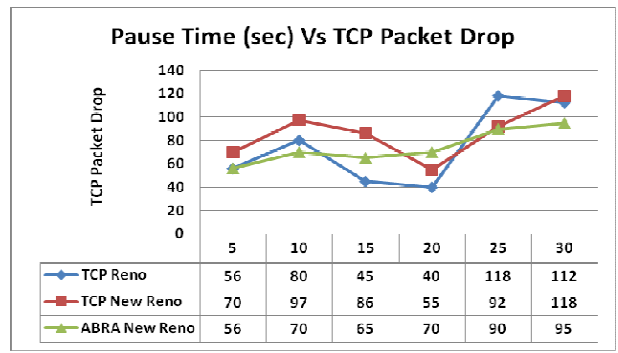

Figure 7

Simulation results in fig. 5, 6 & 7 shows that initially Packets drops are less because of less congestion, less timeout waiting time, low route failure, but upon increments in variants like no. of nodes, node speed and Pause time of the network, packet drop rate is also increases gradually because of Route Failure, high Congestion, destination unreachable, time slice expired, frequently change route and bandwidth, improper routing path and node mobility.

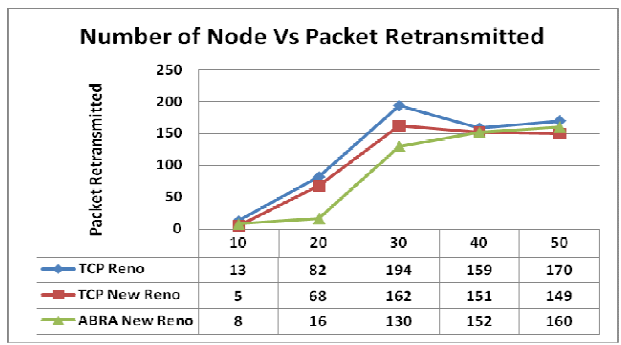

Figure 8

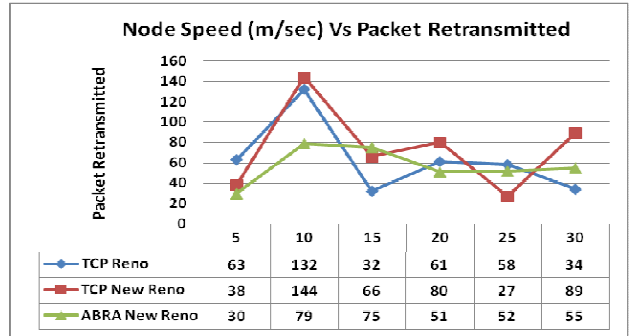
Figure 9

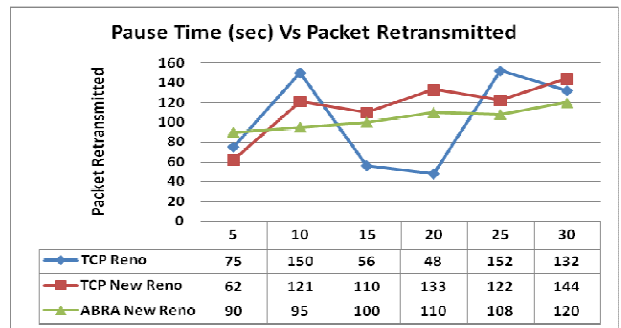
Figure 10

Simulation results in fig. 8-10 shows packet retransmitted upon different variants like variation in number of node speed of node and pause time in network. When number of nodes increases from 10 to 40 in network than packet retransmission is also increases gradually, but Packets retransmission rate vary under increase node mobility and pause time due to increments in packet drop rate, acknowledgment loss, packet delay and also by time out condition.

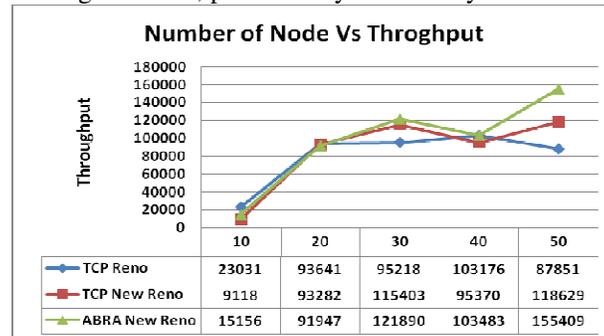
Figure 11

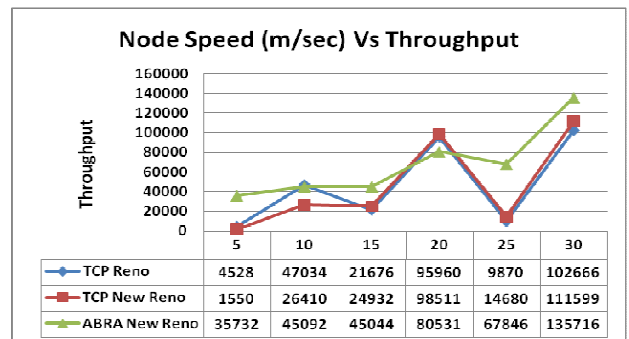
Figure 12

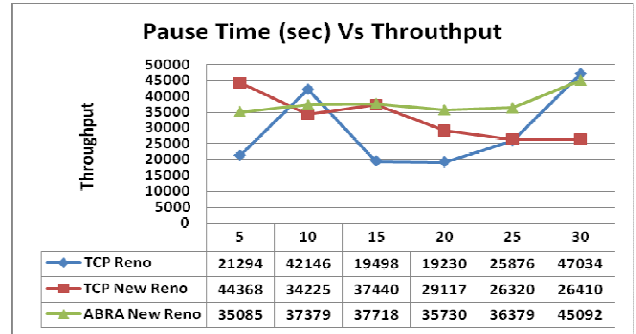

Figure 13

Simulation results in fig. 11-13 shows that throughput of different variants with variation in number of nodes, speed of node and pause time in network. It is observed that throughput of ABRA New Reno TCP is vary according to variation in number of node, speed of node and Pause time. Throughput depends on receive bytes, packet loss and drop acknowledgement. These parameters are inconsistent and greater sometime, because of wrong estimation of bandwidth and delay.

## 8. CONCLUSION

An improve TCP New Reno (ABRA New Reno) is proposed for mobile ad-hoc networks using calculations of New Retransmission Time out, to improve performance in terms of congestion control and implemented in a Mobile Ad-hoc Network under QualNet 4.0 simulator. To calculate New RTO, adaptive backoff response approach was applied in TCP New Reno which suggests ABRA New Reno. Extensive simulation studies were taken to compare its performance with standards TCP Reno and TCP New Reno over Ad-hoc Mobile Network. After implementation ABRA in new Reno it is analysed under varying conditions of node speed, Pause Time and number of node. Simultaneously efficiency of ABRA New Reno on various performance metrics was measured, including data packet received, packet drop, packets retransmitted, throughput.

From results it is observed that, ABRA New Reno performs well under the varying conditions of high density node, high node speed and pause time, because of proper utilization of time, optimal paths between nodes, optimal bandwidth exploitation and less packet delay. Simultaneously throughput was found to be most favourable, however it may be vary according to traffic conditions and congestion.

## 9. FUTURE WORK

Our experimental results suggest that in forthcoming efforts, simulation of ABRA New Reno TCP by varying other parameter and performance in large and realistic scenario will have great potential. Furthermore simulations are to be done in future to explore the effect of ABRA New Reno TCP on other TCP Variants and ad hoc routing protocols. In order to accurately simulate the realistic congested network environment, there is a need to experiment with multiple TCP flows.

International Journal of Wireless & Mobile Networks (IJWMN) Vol. 3, No. 2, April 2011

## Authors

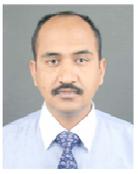
*Sanjeev Sharma received his Doctor of philosophy in Information Technology From RGPV BHOPAL He is currently working with University Teaching Department, RGPV, BHOPAL as department head. He has 20 years of experience in teaching field. His research interest includes network security system, Mobile Ad-hoc network. He has various published national and international papers and journals.*

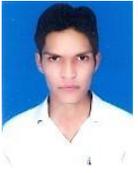
*Dhananjay Bisen received his Bachelor degree in Information Technology from Jabalpur Engg. College Jabalpur in 2007 .He received M.TECH Degree in Information Technology from SOIT. RGPV, Bhopal. Currently he is working with RADHARAMAN INSTITUTE OF TECHNOLOGY & SCIENCE Bhopal. His research Interest Mobile Ad-Hoc Network, Network Security.*